%% file: qcd98.tex
\documentstyle[twoside,fleqn,espcrc2,epsf]{article}

\def\bm#1{{\mbox{\boldmath $#1$}}}

\newcommand\as{{\alpha_s}}

\newcommand\NLO{next-to-leading order }

\newcommand{\Nf}{{N_{\rm f}}}
\newcommand{\yf}{{y_{\rm f}}}

\newcommand{\beq}{\begin{equation}}
\newcommand{\eeq}{\end{equation}}
\newcommand{\beeq}{\begin{eqnarray}}
\newcommand{\eeeq}{\end{eqnarray}}
\newcommand\nn{\nonumber}
\newcommand\bs{\!\!\!\!\!\!\!\!\!\!\!\!}
\newcommand\ycut{y_{\rm cut}}                              
\newcommand\ms{{\overline{{\rm MS}}}}

\title{
Multijet rates in $e^+e^-$ annihilation: perturbation theory versus LEP
data\thanks{This research was supported in part by the EU Fourth Framework
Programme ``Training and Mobility of Researchers'', Network ``Quantum
Chromodynamics and the Deep Structure of Elementary Particles'',
contract FMRX-CT98-0194 (DG 12 - MIHT), as well as by the Hungarian
Scientific Research Fund grant OTKA T-025482 and the Research Group in
Physics of the Hungarian Academy of Sciences, Debrecen.}}
\author{Zolt\'an Nagy
\address{Department of Theoretical Physics, KLTE, H-4010
Debrecen, P.O.Box 5, Hungary}
and Zolt\'an Tr\'ocs\'anyi
\address{Theory Division, CERN, CH-1211 Geneva 23 and\\
Institute of Nuclear Research of the Hungarian Academy of
Sciences, H-4001 Debrecen, P.O.Box 51, Hungary}}

\begin{document}

\begin{titlepage}
\vspace*{-2cm}
\begin{flushright}
CERN-TH/98-266\\
hep-ph/9808364 \\
\end{flushright}
\vskip .5in
\begin{center}
{\large\bf
Multijet rates in $e^+e^-$ annihilation: perturbation theory versus LEP
data\footnote{Presented by Z. Tr\'ocs\'anyi at the International
Euroconference on Quantum Chromodynamics ({\em QCD '98}), Montpellier,
France, July 2--8, 1998}}\\
\vskip 5mm 
Zolt\'an Nagy\\
\vskip 2mm
{\it Department of Theoretical Physics, KLTE, H-4010 Debrecen, P.O.Box 5,
Hungary} \\
\vskip 2mm
 and
\vskip 2mm
Zolt\'an Tr\'ocs\'anyi \\
\vskip 2mm
{\it Theory Division, CERN, CH-1211 Geneva 23} and \\
{\it Institute of Nuclear Research of the Hungarian Academy of
Sciences, H-4001 Debrecen, P.O.Box 51, Hungary} \\
\end{center}
\vskip 15mm 

\begin{center} {\large \bf Abstract} \end{center}
\begin{quote}
We show that the next-to-leading order perturbative prediction, matched
with the next-to-leading logarithmic approximation for predicting both
two-, three- and four-jet rates using the Durham jet-clustering
algorithm, in the $0.001 < \ycut < 0.1$ range gives a very accurate
description of the data obtained at the Large Electron Positron collider.
This information can be utilized either for simultaneous measurement of
the strong coupling and the QCD colour charges, or for improving the
QCD background prediction in new particle searches at LEP2.
\end{quote}

\vfill
\begin{flushleft}
CERN-TH/98-266\\
August 1998 \\
\end{flushleft}
\end{titlepage}
\setcounter{footnote}{0}

\newpage
\begin{abstract}
We show that the next-to-leading order perturbative prediction, matched
with the next-to-leading logarithmic approximation for predicting both
two-, three- and four-jet rates using the Durham jet-clustering
algorithm, in the $0.001 < \ycut < 0.1$ range gives a very accurate
description of the data obtained at the Large Electron Positron Collider.
This information can be utilized either for simultaneous measurement of
the strong coupling and the QCD colour charges, or for improving the
QCD background prediction in new particle searches at LEP2.
\end{abstract}

\maketitle

\section{INTRODUCTION}

Jet production rates provide one of the most intuitive ways to study
the underlying parton structure of hadronic events. However, lacking the
necessary theoretical accuracy, for measuring the strong coupling $\as$,
only the differential $y_3$ distributions were used so far, where $y_3$
is the event shape variable measuring the value of the jet resolution
parameter $\ycut$ for which the jet multiplicity of a given event
changes from three-jet to two-jet. The new \NLO results for four-jet rates
\cite{DSjets,NTjets,Glover} raise the possibility of using LEP four-jet data
for $\as$ precision measurements.

In electron-positron annihilation the widely known Durham \cite{durham}
jet-clustering algorithm has become an indispensable tool for
classifying multihadron final states into jets. This jet algorithm has
the advantages of relatively small hadronization corrections and of the
possibility of resumming large logarithms near the edge of the phase
space (small $\ycut$ region), thus extending the validity of the
perturbative prediction. Recently a new, Cambridge, jet-algorithm
was proposed; it has similar resummation properties, but smaller
hadronization corrections for mean jet multiplicities \cite{cambridge}.
More detailed studies showed, however, that the small hadronization
corrections found for the Cambridge algorithm in the study of the mean
jet rate are due to cancellations among corrections for the individual
jet production rates. Apart from the very small values of the resolution
parameter, $\ycut < 10^{-3.2}$, the Durham clustering shows, for the
individual rates, comparably small (for $\ycut > 10^{-2}$), or even much
smaller, hadronization corrections \cite{cambridgeB}. We will be using
data with $\ycut > 10^{-3}$; therefore, in this talk we consider only
multijet rates obtained using the Durham algorithm.

\section{THE THEORETICAL DESCRIPTION}

The $n$-jet rates are defined as the ratio of the $n$-jet cross section
to the total hadronic cross section, and at \NLO these take the general form
\beeq
\label{nloxsec}
&&
\bs
R_n=\frac{\sigma_{n{\rm -jet}}}{\sigma_{\rm tot}}
= \eta(\mu)^{n-2} B_n(\ycut)
\\ \nn &&\!\!
\bs
+ \eta(\mu)^{n-1}
\Big[(n-2)B_n(\ycut)\beta_0 \ln(x_\mu) + C_n(\ycut)\Big]\:.
\eeeq 
In this equation $\eta(\mu)=\as(\mu) C_F/2\pi$,
$x_\mu=\mu/\sqrt{s}$, where $\mu$ is the renormalization scale and
$\sqrt{s}$ is the total c.m.\ energy. The functions $B_n$ and $C_n$ are
independent of the renormalization scale, $B_n$ is the Born
approximation and $C_n$ is the radiative correction. These functions in
the case of the Durham clustering for $n=2$ and 3 were calculated based
upon the ERT matrix elements \cite{ERT} and for $n=4$ they were
obtained in Refs.~\cite{DSjets,NTjets} based upon the one-loop matrix
elements of Refs.~\cite{loop4parton} and tree-level matrix elements of
Ref.~\cite{NTjets}. We use the two-loop expression
for the running coupling,
\beeq
\label{twoloopas}
&&
\bs
\eta(\mu) =
\\ \nn &&
\frac{\eta(M_Z)}{w(\mu,M_Z)}
\left(
1-\frac{\beta_1}{\beta_0}\eta(M_Z)\frac{\ln(w(\mu,M_Z))}{w(\mu,M_Z)}
\right)\:,
\eeeq
with
\beq
\label{wqq0}
w(q,q_0) = 1 - \beta_0 \eta(q_0)\ln\left(\frac{q_0}{q}\right)\:,
\eeq
\beq
\beta_0 = \frac{11}{3}x - \frac{4}{3} \yf\:,
\eeq
\beq
\beta_1 = \frac{17}{3}x^2 - 2 \yf - \frac{10}{3} x \yf
\eeq
($x = C_A/C_F$ and $\yf = T_R \Nf/C_F = \Nf/2C_F$).
All theoretical predictions in this contribution were obtained for five
light-quark flavours at the $Z^0$ peak with $M_Z = 91.187$\,GeV,
$\Gamma_Z = 2.49$\,GeV, $\sin^2\theta_W=0.23$ and $\as(M_Z) = 0.118$.
%

Multijet fractions decrease very rapidly with increasing resolution
parameter $\ycut$. Consequently, most of the available multijet data are
at small $\ycut$.  It is well known that for small values of $\ycut$
the fixed order perturbative prediction is not reliable, because the
expansion parameter $(\as/2\pi) \ln^2 \ycut$ logarithmically enhances the
higher-order corrections. For instance, $(\as/2\pi) \ln^2 0.01\approx
0.4$. Thus, one has to perform the all-order resummation
of the leading and next-to-leading logarithmic (NLL) contributions.
This resummation is possible for the Durham algorithm using the coherent
branching formalism \cite{durham}. The two-, three- and four-jet rates
in the NLL approximation are given
in terms of the NLL emission probabilities $\Gamma_i(Q,q)$ \cite{durham}
which have the following form:
\beeq
&&
\bs
\Gamma_q(Q,q) =
\\ \nn &&
4\frac{\eta(q)}{q}
\left[\Big(1+\eta(q)\,K\Big)\ln\frac{Q}{q}
-\frac{3}{4}\right]\:,
\\ &&
\bs
\Gamma_g(Q,q) =
\\ \nn &&
4x\frac{\eta(q)}{q}
\left[\Big(1+\eta(q)\,K\Big)\ln\frac{Q}{q}
-\frac{11}{12}\right]\:,
\\ &&
\bs
\Gamma_f(Q,q) =
4\frac{\yf}{3}\frac{\eta(q)}{q}\:.
\eeeq
We relate the $\eta(q)$ strong coupling appearing in the emission
probabilities to the strong coupling at the relevant renormalization
scale, $\eta(\mu)$, according to the one-loop formula
\beq
\eta(q) = \frac{\eta(\mu)}{w(q,\mu)}\:,
\eeq
where $w(q,q_0)$ was defined in Eq.~(\ref{wqq0}),
and we use Eq.~(\ref{twoloopas}) for expressing $\eta(\mu)$ in terms of
$\eta(M_Z)$. We could also use a two-loop formula for $\eta(q)$,
but the result would differ only in subleading logarithms. However, we
take into account a certain part of subleading soft logarithms with the
inclusion of the $K$ term. The $K$ coefficient is renormalization-scheme
dependent. In the $\ms$ scheme it is given by \cite{Kcoeff} 
\beq
K=\left(\frac{67}{18}-\frac{\pi^2}{6}\right)x-\frac{10}{9}\yf\:.
\eeq

The result of this resummation together with its renormalization-scale
dependence in the case of four-jet rates was studied in
Ref.~\cite{NTjets}, where we found that the
fixed-order
and the NLL approximations differ significantly. One expects that for
large values of $\ycut$ the former, and for small values of $\ycut$ the
latter is the reliable description; therefore, the two results have to
be matched.

The Durham multijet rates can be resummed at leading and NLL order,
but they do not satisfy a simple exponentiation \cite{Catani} (except
for the two-jet rate). For an observable that does not exponentiate,
the viable matching scheme is the R-matching \cite{durham}, which we use
according to the following formula:
\beeq
&&
\bs
R_n^{\rm R-match} = R_n^{\rm NLL} +
\\ \nn && 
\Bigg[\eta^{n-2}
\Big(B_n - B_n^{\rm NLL}\Big)
+\eta^{n-1}
\Big(C_n - C_n^{\rm NLL}\Big)\Bigg]\:,
\eeeq
where $B_n^{\rm NLL}$ and $C_n^{\rm NLL}$ are the
coefficients in the expansion of $R_n^{\rm NLL}$
as in Eq.~(\ref{nloxsec}).

\section{RESULTS}

In Figs.~1, 2 and 3, we show the theoretical prediction at the various
levels of approximation: in fixed-order perturbation theory at the Born
level (LO), at \NLO (NLO), resummed and R-matched prediction (NLO+NLL),
and improved resummed and R-matched prediction (NLO+NLL+K), for the
two-, three- and four-jet rates respectively.  Also shown are the
multijet rates measured by the ALEPH collaboration at the $Z^0$ peak
\cite{ALEPHRn} corrected to the parton level using the PYTHIA Monte Carlo
\cite{PYTHIA}. We used bin-by-bin correction and the consistency of the
correction was checked by using the HERWIG Monte Carlo \cite{HERWIG}.
The two programs gave the same correction factors within statistical
errors. The errors on the data are the scaled statistical errors of the
published hadron level data, and we did not include any systematic
experimental error, or the error due to the hadron-to-parton
correction. In the lower parts of the plots we show the relative
difference (data-theory)/theory, where theory means the `NLO+NLL+K'
prediction and we also indicated the renormalization scale dependence.

\begin{figure}
\epsfxsize=7.3cm \epsfbox{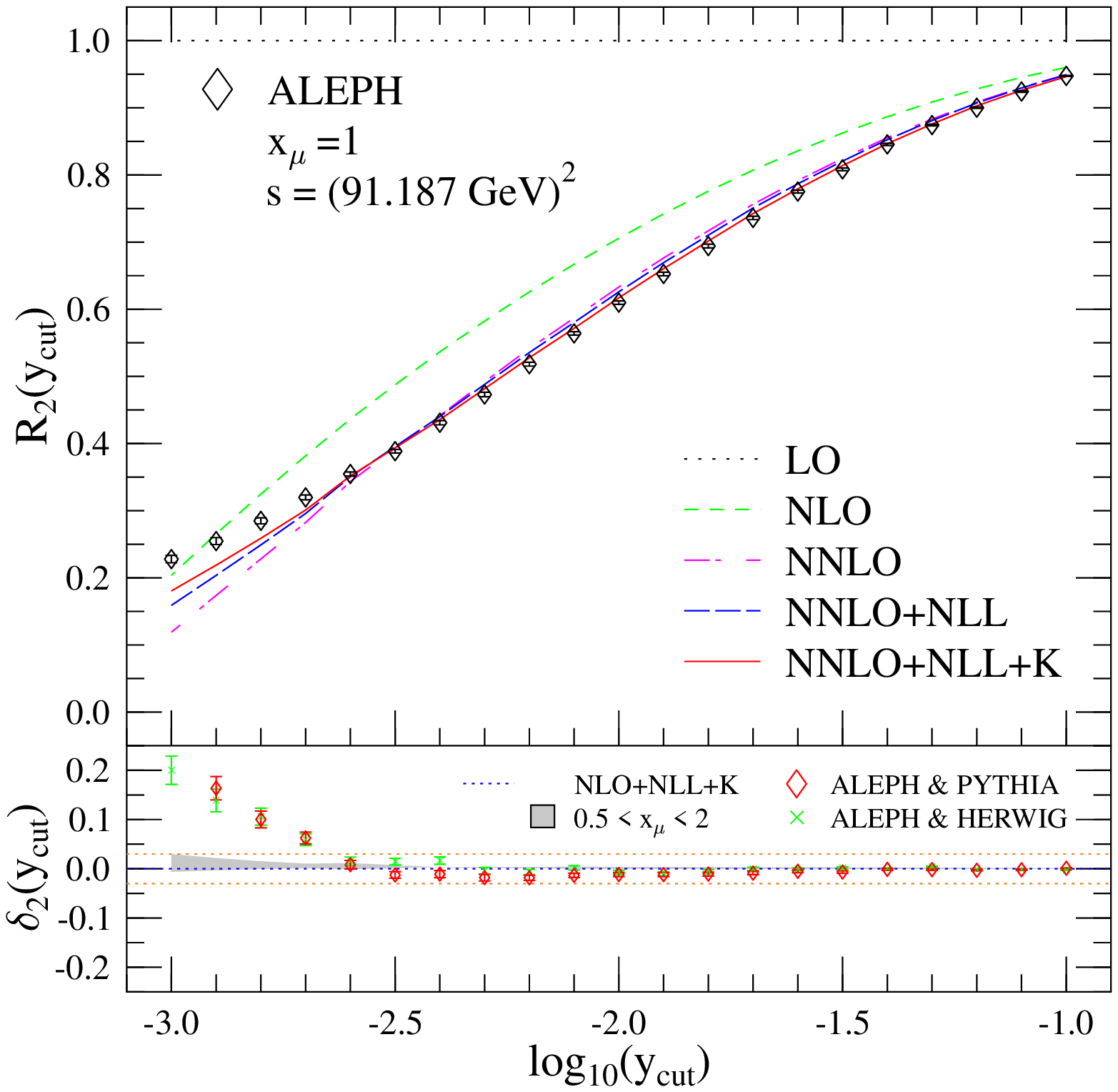} 
\vspace*{-30pt}
\caption{
The QCD predictions for two-jet rates, with renormalization scale is set
to $x_\mu=1$, compared to ALEPH data.
The lower part of the plot shows the relative difference
$\delta=$(data-theory)/theory, where theory means the improved NLL
approximation matched with the NLO result as explained in the text.
The bands indicate the theoretical uncertainty due to the variation of
the renormalization scale $x_\mu$ between 0.5 and 2.
}     
\epsfxsize=7.3cm \epsfbox{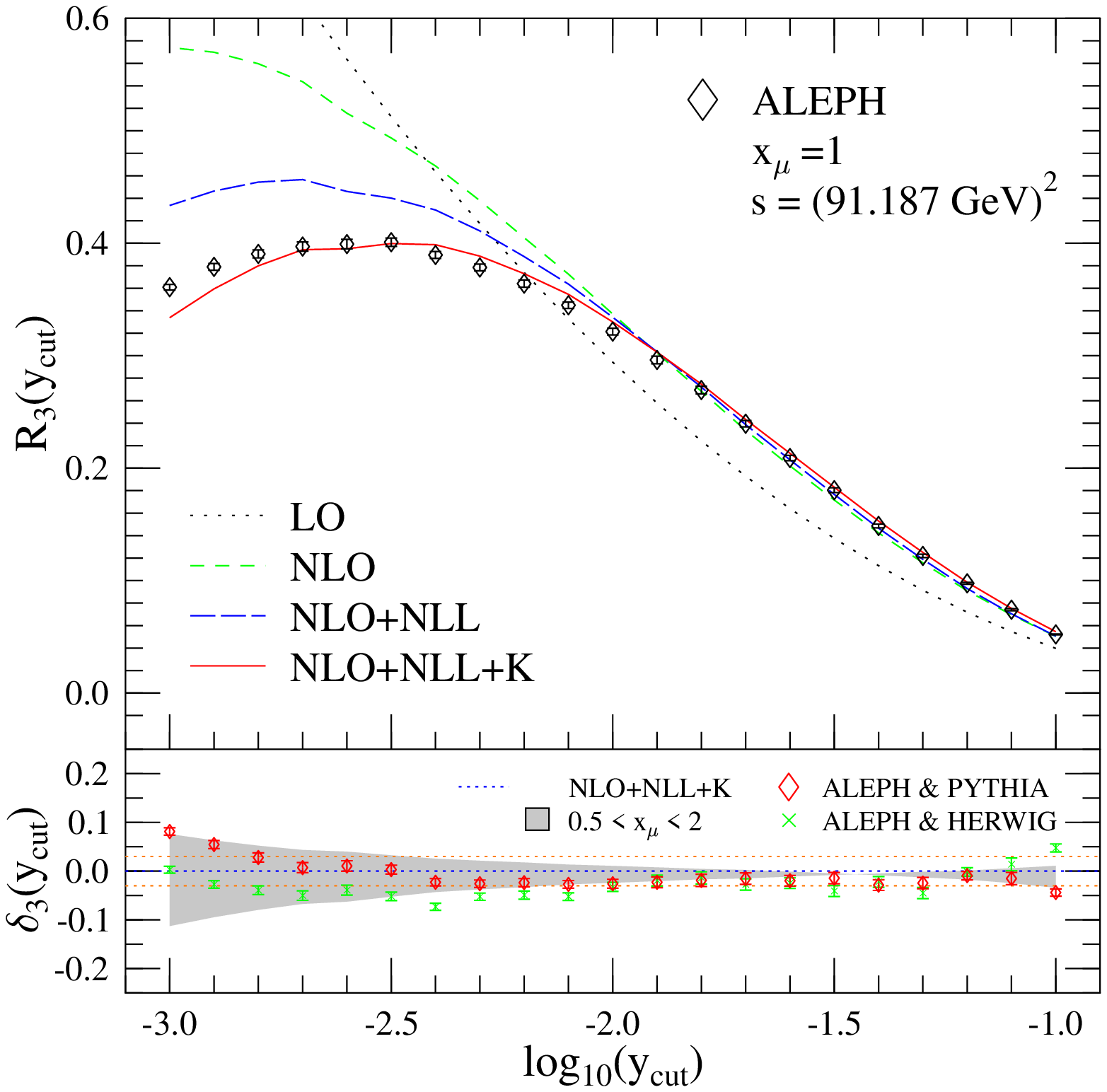} 
\vspace*{-30pt}
\caption{
Same as Fig.~1, but for three-jet rates.
}     
\end{figure}

\begin{figure}[tb]
\epsfxsize=7.3cm \epsfbox{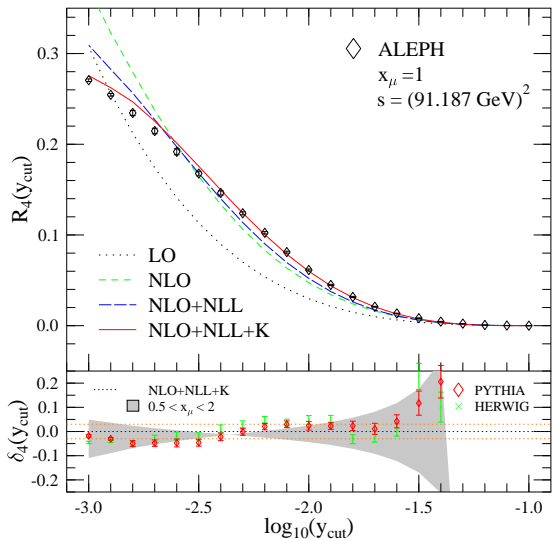} 
\vspace*{-30pt}
\caption{
Same as Fig.~1, but for four-jet rates.
}     
\end{figure}

Figures 1--3 deserve several remarks. First of all, we see that the inclusion
of the radiative corrections improves the fixed-order description of the
data, using the natural scale $x_\mu=1$ for larger values of $\ycut$.
Secondly, the importance of resummation in the small $\ycut$ region is
clearly seen, but it is still not sufficient to describe the data at the
natural scale, neglected subleading terms are still important. On the
other hand, the improved resummation seems to take into account just
the right amount of subleading terms; this makes the agreement between
data and theory almost perfect over the whole $\ycut$ region, as can be
seen from the lower part of the plots. (In the case of four-jet rate,
for $\ycut > 10^{-1.7}$, $\delta_4$ falls outside the $\pm 3\,\%$ band,
one should keep in mind that in this region i) the renormalization
scale dependence is relatively large and ii) the number of events is
very small; therefore the statistical errors of the data and that of
the hadron-to-parton correction are very large.)

We found remarkably small scale dependence for the `NLO+NLL+K'
predictions in the region $\ycut > 10^{-3}$. This feature,
however, should be taken with care. On the one hand at any artificial
narrowing of the scale-dependence bands, e.g.\ at a crossover point,
the bands almost certainly do not represent the size of the truncation error
at that point.  On the other, the improvement, obtained by including
the two-loop coefficient K, affects NNLL terms, but there are other
contributions of the same order that are not taken into account (e.g.\
\NLO running of $\as$ and other dynamical effects).  The scale
dependence of the `NLO+NLL+K' result would consistently be under
control only after the inclusion of the complete set of NNLL terms. One
expects however, that the scale dependence of the `NLO+NLL' prediction
is an upper bound for the scale dependence of the perturbative
prediction, with subleading logarithms taken into account completely.
Therefore, we may use our `NLO+NLL+K' prediction for QCD tests, for
instance, for measuring the strong coupling $\as$ with the condition
that we estimate the systematic theoretical uncertainty due to the
scale dependence from the scale dependence of the `NLO+NLL' prediction
(obtained from varying the scale $x_\mu$ between 0.5 and 2 as standard
choice). The result of such a fit is given in Table 1, where the
central value was obtained using the `NLO+NLL+K' result with $x_\mu=1$
and the error represents
\begin{itemize}
\item the statistical error on the data,
\item the systematic error due to changing the fit range (the whole
range shown in Figs.~2 and 3 was chosen) by one bin at both ends in
both directions, \item the error due to the use of different Monte
Carlo programs (PYTHIA and HERWIG) for calculating the hadronization
corrections,
\item and the error due to the variation of renormalization scale as
described above,
\end{itemize}
all added in quadrature. This error is strongly dominated by the scale
uncertainty.  Of course, we could not include the systematic
experimental error. Also, in Table~1 we show the result of the fit when
the renormalization scale is left as a free parameter.  It is
remarkable that, in the case of the three-jet rate, the natural scale
$x_\mu=1$ is very close to the scale giving the smallest
$\chi^2/$d.o.f.\ ($x_\mu=0.92$). On the other hand, in the case of the
four-jet rate, the fitted scale is still somewhat lower than the
natural scale. In our interpretation, this is due to the importance of
the still neglected subleading terms. Indeed, we could also fit the
four-jet data with the `NLO+NLL' prediction, but with a very low scale
($x_\mu\simeq 0.2$) and a slightly higher value of the strong coupling
($\as(M_Z)=0.121$).

\input Table-I

\section{CONCLUSION}

In this talk we studied the perturbative description of two-, three- and
four-jet rates produced at LEP, obtained using the Durham clustering
algorithm. We found that the best theoretical approximation that is
currently available gives a remarkably precise account of the data. In a
previous publication \cite{NTangulars}, we found that the angular
correlations defined on four-jet events are also well described (within
$\pm$5\,\%) by
perturbative QCD. These observations suggest that the same procedure
should provide an accurate prediction of the multijet backgrounds
encountered in new particle searches and $W$-mass measurements at LEP2.

We also performed a measurement of the strong coupling based upon
multijet rates. We found $\as(M_Z) = 0.1173\pm 0.0018$, where the error
includes the statistical and theoretical ones, but not the systematic
experimental ones.

These results were produced in part by a partonic Monte Carlo program
called DEBRECEN \cite{debrecen}, which is based upon the dipole formalism
\cite{CSdipole} and can be used for the calculation of QCD radiative
corrections to the differential cross section of any kind of infrared-safe
three- and four-jet observable in electron-positron annihilation.

\def\ibid#1#2#3  {{\it ibid} {\bf #1}, #2 (19#3)}
\def\np#1#2#3  {Nucl.\ Phys.\ #1 (19#3) #2}
\def\pl#1#2#3  {Phys.\ Lett.\ #1 (19#3) #2}
\def\plb#1#2#3  {Phys.\ Lett.\ B #1, #2 (19#3)}
\def\prep#1#2#3  {Phys.\ Rep.\ #1 (19#3) #2}
\def\prd#1#2#3 {Phys.\ Rev.\ D #1 (19#3) #2}
\def\prl#1#2#3 {Phys.\ Rev.\ Lett.\ #1 (19#3) #2}
\def\zpc#1#2#3  {Zeit.\ Phys.\ C #1 (19#3) #2}
\def\epc#1#2#3  {Eur.\ Phys.\ C #1 (19#3) #2}
\def\cpc#1#2#3  {Comput.\ Phys.\ Commun.\ #1 (19#3) #2}
\def\jhep#1#2#3 {JHEP #1, (19#3) #2}
\def\etal{et al.}

\end{document}

%% file: Table-I.tex
\begin{table}             
\caption{Results of $\as(M_Z)$ fits with fixed and fitted renormalization
scale using multijet data obtained by the ALEPH collaboration at LEP1.}
\begin{tabular}{lccc}
  \hline
  \hline
                 & $R_3$          &  $R_4$      & $R_3$ \& $R_4$     \\
  \hline
$\bm{\as(M_Z)}$  & 0.116          & 0.1182      & 0.1175             \\
$x_\mu$ fixed    &    1           & 1           & 1                  \\
$\chi^2/$d.o.f.\ & 5.9/17         & 29.2/12     & 54.4/29            \\
  \hline
errors           &                &             &                    \\
  \hline
statistical      & $\pm$0.0004    & $\pm$0.0003 & $\pm$0.0002        \\
fit range        & $\pm$0.0002    & $\pm$0.0002 & $\pm$0.0001        \\
hadroniz.\       & $\pm$0.0008    & $\pm$0.0005 & $\pm$0.0001        \\
ren.\ scale      & $\pm$0.0015    & $\pm$0.0022 & $\pm$0.0017        \\
  \hline
total            & $\pm$0.0018    & $\pm$0.0023 & $\pm$0.0018        \\
  \hline
$\bm{\as(M_Z)}$  & 0.116          & 0.1176      & 0.1173             \\
$x_\mu$ fitted   & 0.92           & 0.64        & 0.7                \\
$\chi^2/$d.o.f.\ & 5.5/17         & 10.4/12     & 29.6/29            \\
  \hline                        
  \hline                        
\end{tabular} 
\end{table}
\noindent